\documentclass[12pt]{article}
\usepackage[a4paper, hmargin=2.5cm, vmargin=2.5cm]{geometry}
\usepackage{amsfonts,amsmath,graphicx,multirow,xcolor}
\newcommand{\bPsi}{\mbox{\boldmath $\Psi$}}

\begin{document}
\title{Mining Financial Data using Mixtures of Mirrored Weibull Distributions}
\author{Zijun Jia$^{1, \star}$, Sharon X. Lee$^{2}$}
\date{}

\maketitle

\begin{flushleft}
$^1$Science Department, Inner Mongolia Agricultural University, Hohhot, China .\\
$^1$School of Mathematics and Physics, University of Queensland, Brisbane, Australia.\\
$^\star$ E-mail: jzj1018@imau.edu.cn 
\end{flushleft}

\begin{abstract}

Risk management is an important part of financial practice, essential for protecting assets and investments in modern-day volatile markets. This paper proposes a mixture of mirrored Weibull (MMW) distribution for modelling stock returns and estimating risk measures. Unlike common practices which are typically based on the normal distribution, the MMW model can flexibly accommodate non-normal features frequently exhibited in financial data. It also enjoys appealing properties such as having a simple density expression and fast parameter estimation.  We demonstrate the effectiveness of our model by assessing its performance in Value-at-Risk (VaR) estimation of three S\&P500 stocks. The MMW model compares favourably to Gaussian mixture model and $t$-mixture model, with significant improvements in VaR estimation and prediction. 
  
\end{abstract}

\section{\large Introduction}
\label{s:intro}

In the ever-evolving landscape of financial markets, effective risk management is critical for both institutional and retail investors. The dynamic nature of financial markets necessitates a rigorous approach to risk management, particularly in the context of asset pricing and investment strategies.  
To this end, various risk assessment methods have been proposed, each with their own merits. Among these, Value-at-Risk (VaR) is one of the most widely used metrics. This statistical measure quantifies the potential loss in value of an investment portfolio over a specified time period at a predefined confidence level. However, the estimation of VaR can be challenging, especially in finding models that can accurately capture the underlying distributions of asset returns.         

Most existing VaR estimation methods relies on some distributional assumptions on the financial data, typically the normality assumption. However, ample empirical studies have suggested that real financial returns rarely follow the normal distribution. Instead, these data frequently exhibit non-normal distributional features such as heavy tails, skewness, kurtosis, and sometimes, multimodality. Traditional methods that rely on these normality and/or linearity assumptions can lead to misleading results when applied to non-normal return distributions. In view of this, new approaches have been proposed, incorporating more sophisticated techniques or adopting more flexible models.  
In recent years, mixture models have emerged as a more effective alternative to conventional methods for estimating VaR \cite{b1, b2, b3, b4}. Well-studied in the Statistics literature, mixture models offer a flexible framework for modelling heterogeneous data and can accommodate the diverse characteristics of asset returns. 
Several studies have demonstrated the effectiveness of Gaussian mixture model (GMM) over traditional VaR estimation methods, offering more reliable estimates by more accurately capturing the non-normal characteristics of return distributions \cite{b8, b5, b6, b13, b7, b9}. A drawback of the GMM approach is that it may require multiple Gaussian components to accommodate a non-normal cluster, thus increasing model complexity and computation time. An alternative is to employ non-normal component distributions (such as skew distributions) that can inherently capture non-normal features \cite{b14, b11, b12}. While these models have shown considerable improvements in VaR estimation accuracy, they also come with higher computational costs.

We propose a mixture of mirrored Weibull (MMW) distributions for modelling financial returns. The Weibull distribution has the advantages of having a simple analytic form and can accommodate skewness and heavy-tailedness. It thus represents a good compromise between GMM and skew mixture models, with improved performance over GMM and lower computation time than skew mixture models. However, the Weibull distribution cannot be directly applied to financial return data as the distribution is defined only on the non-negative domain. To overcome this limitation, we propose the mirrored Weibull distribution, a transformation of the Weibull distribution that retains the simplicity of its density expression. Parameter estimation of the MMW model can be carried out using maximum likelihood via the Expectation-Maximization (EM) algorithm. For fully unsupervised learning, the number of components is selected based on the Bayesian information criterion (BIC). We demonstrate the effectiveness of our method using three stocks listed in the S\&P500 index.        

The rest part of paper is organized as follows. We introduce the MMW model in Section \ref{sec:MMW} and presents a parameter estimation algorithm for it in Section \ref{sec:EM}. Section \ref{sec:VaR} describe the VaR estimation, forecast, and performance evaluation procedure. An application to three S\&P500 stocks is discussed in Section \ref{sec:data} and the performance of the MMW model is compared to GMM and the $t$ mixture model ($t$MM). Finally, concluding remarks are given in Section \ref{sec:concl}.  

\section{Mixtures of Mirrored Weibull (MMW) Distributions}
\label{sec:MMW}

The Weibull distribution \cite{b15} is a versatile probability distribution widely used in various fields including engineering, energy, life data analysis, and social sciences. It is particularly notable for its exponential and extreme value characteristics, and thus is well-suited for modelling heavy-tailed time series data and the occurrence of extreme events \cite{b16}. 
The density of the (two-parameter) Weibull distribution can be expressed as
\begin{equation}
f_{\text{W}}(x; \mu, \sigma) = \frac{\sigma}{\mu} \left(\frac{x}{\mu}\right)^{(\sigma-1)} e^{-\left(\frac{x}{\mu}\right)^{\sigma}}, \label{eq:weibull}
\end{equation}
for $x\geq 0$, where $\mu>0$ is the scale parameter and $\sigma>0$ is the shape parameter. The Weibull distribution exhibits light tail when $\sigma \leq 1$, and heavy tail when $\sigma > 1$. The same parameter also regulates the skewness of the distribution -- symmetric when $\sigma=1$ and skewed otherwise. 

As noted above, the Weibull distribution is prohibitive for modelling financial return data as the density (\ref{eq:weibull}) is defined for $x\geq 0$ only. A simple adjustment to circumvent this is to introduce a shift parameter so that the distribution is defined over the shifted range. However, parameter estimation for this three-parameter Weibull distribution can be difficult \cite{b17}. A generalised Weibull distribution was proposed in \cite{b18} which is a two-sided version of the Weibull distribution that is defined over the entire real line. It has four parameters, a scale parameter and a shape parameter for each side. As can be expected, its density is a little more sophisticated and so is parameter estimation for this model. 

\begin{figure}[htbp]
\centering
\includegraphics[width=0.8\textwidth]{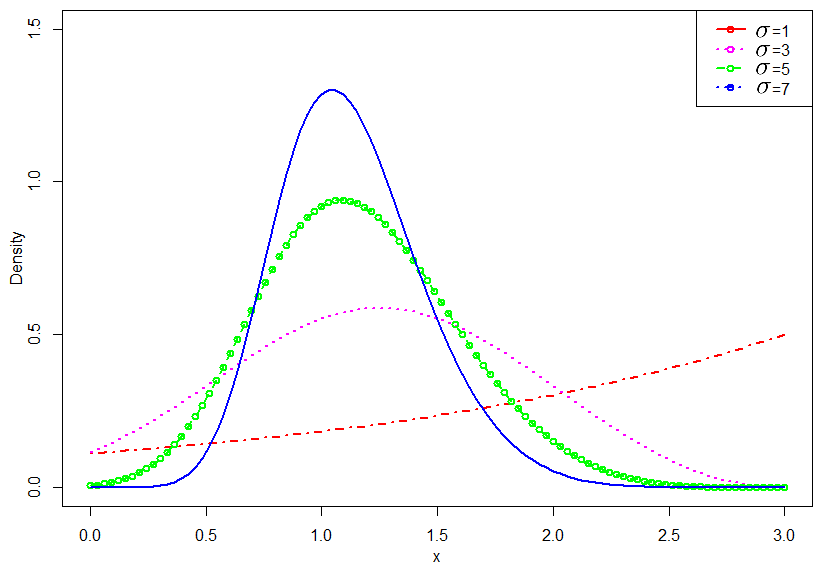}
\caption{Density of mirrored Weibull distributions with different $\sigma$ parameters.}
\label{fig:MW}
\end{figure}

We propose a more simple yet useful modification of (\ref{eq:weibull}), namely the mirrored Weibull distribution. It can be viewed as a special case of an affine transformation of the Weibull distribution that resembles a shifted reflection of (\ref{eq:weibull}), with density given by
\begin{equation}
f_{\text{MW}}(x_j; \mu, \sigma) = \frac{\sigma}{\mu} \left(\frac{c-x_j}{\mu}\right)^{(\sigma-1)} e^{-\left(\frac{c-x}{\mu}\right)^{\sigma}}, \label{eq:MW}
\end{equation}   
for $x \leq c$, where $c = \left\lceil |x_{(n)}|\right\rceil$ and $x_{(n)}$ is the largest observation in the sample. Note that the mirrored Weibull distribution is a two-parameter distribution as $c$ is not considered as a parameter. It is important to note that (\ref{eq:MW}) is different to the three-parameter or shifted Weibull distribution, although their density appear similar. In particular, the shifted Weibull distribution is defined on $x \geq c$ whereas (\ref{eq:MW}) is defined on $x \leq c$. We find in our experiments that the mirrored Weibull distribution usually provides a closer fit to the stock returns data in Section \ref{sec:data}. Examples of the mirrored Weibull distribution are shown in Figure \ref{fig:MW}. Here the scale parameter is fixed at $\mu=1$ and the shape parameter ranges from $\sigma=1$ to $\sigma=7$ (the typical range for $\sigma$ as observed in the experiments in Section \ref{sec:data}).            

We propose a (finite) mixture model based on the mirrored Weibull distribution. The mixtures of mirrored Weibull (MMW) distributions has density given by
\begin{equation}
f(x; \bPsi)=\sum_{i=1}^{g} \pi_i f_{\text{MW}}(x;\mu_i, \sigma_i), 
\label{eq:MMW}
\end{equation}
where $f_{MW}(x;\cdot)$ refers to (\ref{eq:MW}), $g$ is the number of components of the mixture model, $\pi_i$ is the $i$th component weight, and $\Psi$ is the vector of unknown parameters of the model. The component weights, known as the mixing proportions, satisfy the non-negativity and normalisation constraints; that is, $\pi _i\geq 0$ for $i=1,2,\dots,g$ and $\sum_{i=1}^{g} \pi _i=1$. The vector $\Psi$ is given by $\bPsi =(\pi_1,\dots ,\pi_{g-1},\mu_1, \dots, \mu_g, \sigma_1, \ldots, \sigma_g)$. Accordingly, the $g$-component MMW model has $m=3g-1$ free parameters, the same as a $g$-component GMM. However, it has more flexible distributional shapes compared to the GMM.

\section{Parameter Estimation for the MMW Model}
\label{sec:EM}

The mixtures of mirrored Weibull distribution can be fitted using maximum likelihood. The usual approach for fitting mixture model is via the Expectation-Maximization (EM) algorithm \cite{b19}. It is an iterative procedure that begins with an initialization step, then loops between the E- and M-steps until a specified stopping criterion is met. The criterion is typically based on the log-likelihood function, which, for MMW is given by 
\begin{equation}
l(\bPsi) = \sum_{j=1}^n \log f_{\text{MW}}(x_j; \bPsi)
\label{eq:logL}
\end{equation}
for a random sample $X_1, X_2, \ldots, X_n$. 
We now describe the steps of the EM-algorithm for the MMW model. Note that the algorithm assumes $g$ (the number of mixture components) to be fixed and known. We will discuss how $g$ is chosen at the end of this Section.

\subsection{Initialisation}
To begin, we apply the `mirror' transformation to the data: set $y_j = c - x_j$ for $j=1,2,\ldots, n$, where $c$ is defined in (\ref{eq:MW}). An initial clustering of the data can be obtained using $k$-means, hierarchical clustering, random allocation, or similar methods. 
Under the EM framework, a set of binary variables $Z_{ij}$ $(i=1,2,\ldots, g; j=1,2, \ldots, n)$ are introduced to represent the component membership of each observation. Formally, set $z_{ij}^{(0)}=1$ if observation $y_j$ is assigned to cluster $i$ according to the initial clustering, and 0 otherwise. Here we use the superscript $(k)$ to denote the current iteration number of the algorithm (for example, $k=0$ represent the initialisation step). 
Then an initial estimate of $\pi_i$ is given by 
$$\pi_i^{(0)} = \sum_{j=1}^n z_{ij}^{(0)},
$$
and the parameters $\mu_i$ and $\sigma_i$ can be estimated using the method of moments. For $\sigma_i^{(0)}$, it is obtained by solving for $\sigma_i$ numerically the equation  
$$ \frac{s_i^2}{\bar{y}_i^2} = \frac{\Gamma\left(1+\frac{2}{\sigma_i}\right)-\Gamma\left(1+\frac{1}{\sigma_i}\right)^2}{\Gamma\left(1+\frac{1}{\sigma_i}\right)^2},$$
where $\overline{y}_i=\frac{\sum_{j=1}^{n}z^{(0)}_{ij} y_j}{\sum_{j=1}^{n}z^{(0)}_{ij}} $ and $s_i^2=\frac{\sum_{j=1}^{n}z^{(0)}_{ij}(y_j-\overline{y_i})^2}{\left(\sum_{j=1}^{n}z^{(0)}_{ij}y_j\right)-1}  $ is the sample mean and the sample variance, respectively, of the $i$th cluster.  
For $\mu_i^{(0)}$, is is given by 
$$\mu_i^{(0)} = \bar{y}_i \left[\Gamma\left(1+\textstyle\frac{1}{\sigma_i^{(0)}}\right)\right]^{-1}.$$
The initial log likelihood $l^{(0)} = l\left(\bPsi^{(0)}\right)$ is calculated using (\ref{eq:logL}), evaluated at $\bPsi^{(0)}$, given by
\begin{align}
l^{(0)} &= \sum_{i=1}^{g} \sum_{j=1}^{n} \left\{ z_{ij}^{(0)}\log \pi_i^{(0)}
  +z_{ij}^{(0)} \left[\log\mu_i^{(0)}-\mu_i^{(0)}\log\sigma_i^{(0)}
\right.\right.\nonumber\\ 
&\quad\left.\left. + (\mu_i^{(0)}-1)\log y_j-\left(\textstyle\frac{y_j}{\sigma_i^{(0)}}\right)^{\mu_i^{(0)}}\right]\right\}.\nonumber
\end{align} 
This completes the initialisation step. The iteration counter $k$ is incremented by one.  

\subsection{E-step}
At the E-step of the $k$th iteration of the EM algorithm, we update the posterior probabilities (responsibilities) of each observation $y_j (j=1,2,\dots,n)$ belonging to each component $i (i=1,2,\ldots, g)$. This is given by the conditional expectation of $Z_{ij}$ given the current estimate of $\bPsi^{(k)}$, that is, 
\begin{equation}
z_{ij}^{(k)}=\frac{\pi_i^{(k)} f_{\text{W}}\left(y_j;\mu_i^{(k)},\sigma_i^{(k)}\right)}{\sum_{h=1}^{g} \pi_h^{(k)} f_{\text{W}}\left(y_j;\mu_h^{(k)},\sigma_h^{(k)}\right)}. 
\nonumber
\end{equation}

\subsection{M-step}
During the M-step, the parameters of MMW model are updated by maximising the so-called $Q$-function, given by  
\begin{equation}
Q(\bPsi, \bPsi^{(k)})=\sum_{i=1}^{g} \sum_{j=1}^{n} z_{ij}^{(k)} \log(\pi_i f_{W}(y_j;\mu_i,\sigma_i )),\nonumber
\end{equation}
leading to the following equations
\begin{eqnarray}
\pi_i^{(k)} &=& \frac{1}{n} \sum_{j=1}^{n} z_{ij}^{(k)},
\label{eq:pi} \\
\mu_i^{(k)} &=& \left[\frac{\sum_{j=1}^{n} z_{ij}^{(k)} y_j^{\mu_i^{(k)}}\log y_j}
    {\sum_{j=1}^{n} z_{ij}^{(k)} y_j^{\mu_i^{(k)}}} -\frac{\sum_{j=1}^{n} z_{ij}^{(k)}\log y_j}{\sum_{j=1}^{n} z_{ij}^{(k)}} \right]^{-1},
\label{eq:mu}\\
\sigma_i^{(k)} &=& \left[\frac{\sum_{j=1}^{n} z_{ij}^{(k)} y_j^{\mu_i^{(k)}}}
    {\sum_{j=1}^{n} z_{ij}^{(k)}}\right]^{\frac{1}{\mu_i^{(k)}} }.
\label{eq:sigma}
\end{eqnarray}
Note that there is no analytical solution to (\ref{eq:mu}) for $\mu_i^{(k)}$. In practice, (\ref{eq:mu}) is solved using numerical procedures. If computation time is a concern, one can adopt the modified approach of \cite{b20} which provides an approximation for $\mu_i^{(k)}$ without the need of solving an equation numerically. 

\subsection{Stopping criterion}

After each M-step, we check whether the stopping criterion is met. If so, the algorithm terminates and $\bPsi^{(k)}$ is taken to be the final estimate for $\bPsi$. Otherwise, $k$ is incremented by one and we return to the E-step. 
The stopping criterion is typically based on a convergence condition and the pre-specified maximum number of iterations $(k_{\text{max}})$. 

For the experiment in Section \ref{sec:data}, we terminate the algorithm if $k \geq k_{\text{max}}=500$ or if the absolute difference between the log likelihood value of the current and the previous iteration is below the threshold $\epsilon = 10^{-6}$, that is, 
\begin{equation}
\left |l^{(k)}-l^{(k-1)}  \right | < \varepsilon, \nonumber
\end{equation}
where   
\begin{align}
l^{(k)} &= \sum_{i=1}^{g} \sum_{j=1}^{n} \left\{ z_{ij}^{(k)}\log \pi_i^{(k)}
  +z_{ij}^{(k)} \left[\log\mu_i^{(k)}-\mu_i^{(k)}\log\sigma_i^{(k)}
\right.\right.\nonumber\\ 
&\quad\left.\left. + (\mu_i^{(k)}-1)\log y_j-\left(\textstyle\frac{y_j}{\sigma_i^{(k)}}\right)^{\mu_i^{(k)}}\right]\right\}.\nonumber
\end{align} 

\subsection{Choosing $g$}

To determine the number of components $g$, it is common practice to run the EM algorithm for a range of values of $g$ and then perform model selection. In this study, we considered $g=1,2,3,4$ and select the model with the lowest Bayesian information criterion (BIC) value. The BIC for the MMW model is given by $\text{BIC} = m \log n - 2 \log \hat{l}$, where $\hat{l}$ is the log-likelihood value at the final iteration of the EM algorithm.  
Based on the experiment in Section \ref{sec:data}, we found that the most frequently chosen value for $g$ is $2$, suggesting a two-component MMW model may suffice for typical use.   

\section{Value-at-Risk (VaR)}
\label{sec:VaR}

Mathematically, Value-at-Risk (VaR) is defined as the quantile of the loss distribution of the investment portfolio. Let $X$ be the loss distribution. For a given confidence level $\alpha$ $(\alpha \in (0, 1))$, the $\alpha$-level VaR is the $(1-\alpha)$th quantile of $X$, that is, 
$$\text{VaR}_\alpha = F_X^{-1}(1-\alpha),$$  
where $F_X(\cdot)$ is the cumulative distribution function (cdf) of $X$ and $F^{-1}_X(\cdot)$ is the corresponding quantile function. 
For stock returns, we are typically interested in the 1\% VaR and sometimes also the $5\%$ VaR.  

\subsection{VaR Estimation}

It follows from the definition of VaR that an estimate of it can be provided by the corresponding quantile function of the fitted model. 
Although the quantile function of the mirrored Weibull distribution have a simple analytic expression, the case of mixture model is much more difficult. However, the cdf of the MMW model can be written in simple closed form and can be used to facilitate the calculation of quantile values. Nevertheless, it is often quite easy and computationally efficient to find approximations of the quantile values via simulation. Hence, we will employ the simulation approach to obtain VaR estimates based on a fitted model. 

\subsection{One-day VaR Forecasts}

To produce one-day ahead forecasts for VaR, the model is trained using the previous 250 days of return data, which corresponds to approximately one year of historical data. The predicted VaR is then compared to the VaR calculated using the historical method. This procedure is then repeated for next day, with the 250-day window advancing by one-day. The process continues until we reach the the last available date of the dataset, keeping the same rolling window size.  

\subsection{Assessments of VaR Estimates}

We will use two of the most commonly used statistical tests to assess the quality of VaR estimates/forecasts produced by the models. 
Kupiec's test \cite{b21} (also known as the proportion of failures test, the unconditional coverage test, or the backtest) evaluates whether the predicted VaR accurately captures the proportion of actual losses that exceed the VaR threshold. A well-calibrated VaR model should yield an observed number of exceedances that closely align with the expected proportion of exceedance. Failing this test  suggests that the VaR model may not be accurate. 
Christoffersen's test \cite{b22}, also known as the independence test, examines the independence of exceedances over time. It assesses whether the VaR breaches are independent events or if they are correlated. Failing this test may suggests that the model cannot accurately capture the underlying marketing dynamics and thus its estimates may not be as reliable.    

\section{Application to S\&P500 stocks}
\label{sec:data}

In this section, we assess the VaR estimates based on the proposed mixtures of mirrored Weibull (MMW) distributions and compared them with those based on the Gaussian mixture model (GMM) and the $t$-mixture model ($t$MM), using stocks listed on the United States S\&P500 index.   

\subsection{The data}

The data consist of three stocks: the monthly returns for the shares Berkshire Hathaway (BRK) and Walmart (WMT) from June 1996 to May 2024, and the daily returns for the share CVS Health Corporation (CVS) from July 2019 to Jun 2024. The daily/monthly returns of each share were calculated on the adjusted closing prices, and the results were recorded as a log price ratio. Summary statistics (Table \ref{tab:stocks}) indicates mild skewness and moderate to high kurtosis.  

\begin{table}[htbp]
\caption{Summary statistics of the monthly ($^*$daily) returns of three United States S\&P500 stocks for the period of mid-1997 ($^*$mid-2009) to 2024}
\begin{center}
\begin{tabular}{|c|rr|r|}
\hline
	& \textbf{BRK} & \textbf{WMT} & \textbf{CVS$^*$} \\
\hline
minimum & -15.39  & -23.24 & -10.34\\
maximum & 23.42 & 23.58 & 18.44 \\
mean & 0.89 & 0.93 & 0.00 \\
std. dev. & 5.44 & 6.01 & 1.85 \\
skewness & -0.01 & -0.16 & 1.32 \\
kurtosis & 4.22 & 4.45 & 17.36 \\
\hline
\end{tabular}
\label{tab:stocks}
\end{center}
\end{table}

\subsection{Results on BRK}

\begin{figure}[htbp]
\centering
\includegraphics[width=0.8\textwidth]{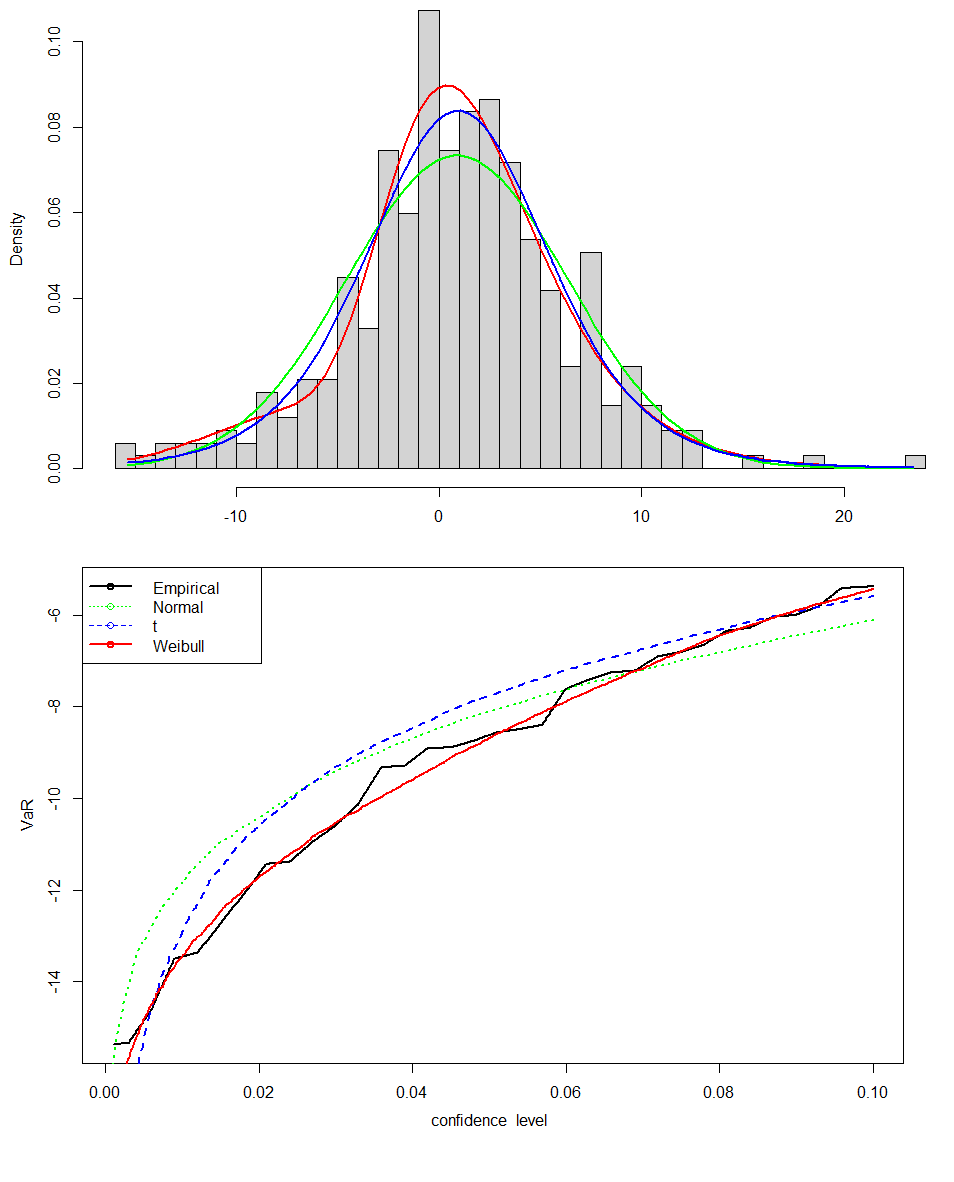}
\caption{(Top) Fitted Gaussian mixture model (green), $t$-mixture model (blue), and mirrored Weibull mixture model (red) to the histogram of BRK returns. (Bottom) The estimated VaR based on the three mixture models and historical method (black).}
\label{fig:BRK}
\end{figure}

The VaR models were trained using the full 7 years of historical monthly log returns. Visual inspection of the fitted models (Figure \ref{fig:BRK} (top)) suggests the models are somewhat similar to each other, with the MMW model (red curve) providing a slightly closer fit to the data. The differences are appreciably wider when we focus on the left tail of the densities, which is the region of interest in practice. The VaR estimates based on MMW is markedly closer to the empirical VaR values for almost the whole interval of 0\% to 10\% confidence level (see Figure \ref{fig:BRK} (bottom)). This is supported by numerical results in Table \ref{tab:BRK}, where the VaR estimate by MMW is closest to the empirical value of -13.46 and -8.66 at $\alpha=5\%$ and $\alpha=1\%$ respectively. It also passes both the Kupiec test and the independence test, whereas both the GMM and the $t$MM models failed the independence test at 1\% confidence level (shown in red in Table \ref{tab:BRK}). 

\begin{table}[htbp]
\caption{Assessments of VaR estimates for BRK given by three different mixture models (Blue = pass; red = fail; bold = best).}
\begin{center}
\begin{tabular}{|c|c|c|c|c|c|c|}
\hline 
Confidence & \multirow{2}*{Model} & \multirow{2}*{VaR} & \multicolumn{2}{|c|}{Kupiec test} & \multicolumn{2}{|c|}{Independence test}\\ 
\cline{4-7} 
level ($\alpha$) & & & LR$_\text{POF}$ & $p$-value & LR$_\text{IND}$ & $p$-value  \\
\hline
\hline
\multirow{3}*{5\%} & GMM & -11.74 & 3.06 & \color{blue}{0.08} & 0.30 & \color{blue}{0.58} \\
\cline{2-7} 
& $t$MM & -12.99 & 0.71 & \color{blue}{0.40} & 0.15 & \color{blue}{0.70} \\
\cline{2-7}
& MMW & \color{blue}{\textbf{-13.43}} & 0.12 & \color{blue}{0.73} & 0.10 & \color{blue}{0.76} \\
\hline
\hline
\multirow{3}*{1\%} & GMM & -8.08 & 0.63 & \color{blue}{0.43} & 8.41 & \color{red}{0.00} \\
\cline{2-7}
& $t$MM & -7.81 & 0.63 & \color{blue}{0.43} & 8.41 & \color{red}{0.00} \\
\cline{2-7}
& MMW & \color{blue}{\textbf{-8.66}} & 0.00 & \color{blue}{0.95} & 3.81 & \color{blue}{0.05} \\
\hline
\end{tabular}
\label{tab:BRK}
\end{center}
\end{table}

\subsection{Results on WMT}

Similar results were observed for the WMT data, where the MMW model provides a closer fit to the data than GMM and $t$MM, especially in the left tail region. At 1\% confidence level, the MMW model outperforms the GMM and the MMW models in all assessments (Table \ref{tab:WMT}; empirical VaR = -16.31). 

\begin{table}[htbp]
\caption{Assessments of VaR$_{0.01}$ estimate for WMT given by three different mixture models (Blue = pass; bold = best).}
\begin{center}
\begin{tabular}{|c|c|c|c|c|c|}
\hline
\multirow{2}*{Model} & \multirow{2}*{VaR$_{0.01}$} & \multicolumn{2}{|c|}{Kupiec test} & \multicolumn{2}{|c|}{Independence test}\\ 
\cline{3-6} 
& & LR$_\text{POF}$ & $p$-value & LR$_\text{IND}$ & $p$-value  \\
\hline
GMM & -13.08 & 3.06 & \color{blue}{0.08} & 0.30 & \color{blue}{0.58}\\
\hline
$t$MM & -14.32 & 1.71 & \color{blue}{0.19} & 0.22 & \color{blue}{0.64} \\
\hline
MMW & \color{blue}{\textbf{-16.26}} & 0.12 & \color{blue}{0.73} & 0.10 & \color{blue}{0.76}\\
\hline
\end{tabular}
\label{tab:WMT}
\end{center}
\end{table}

\subsection{Results on CVS}

\begin{figure}[htbp]
\centering
\includegraphics[width=0.8\textwidth]{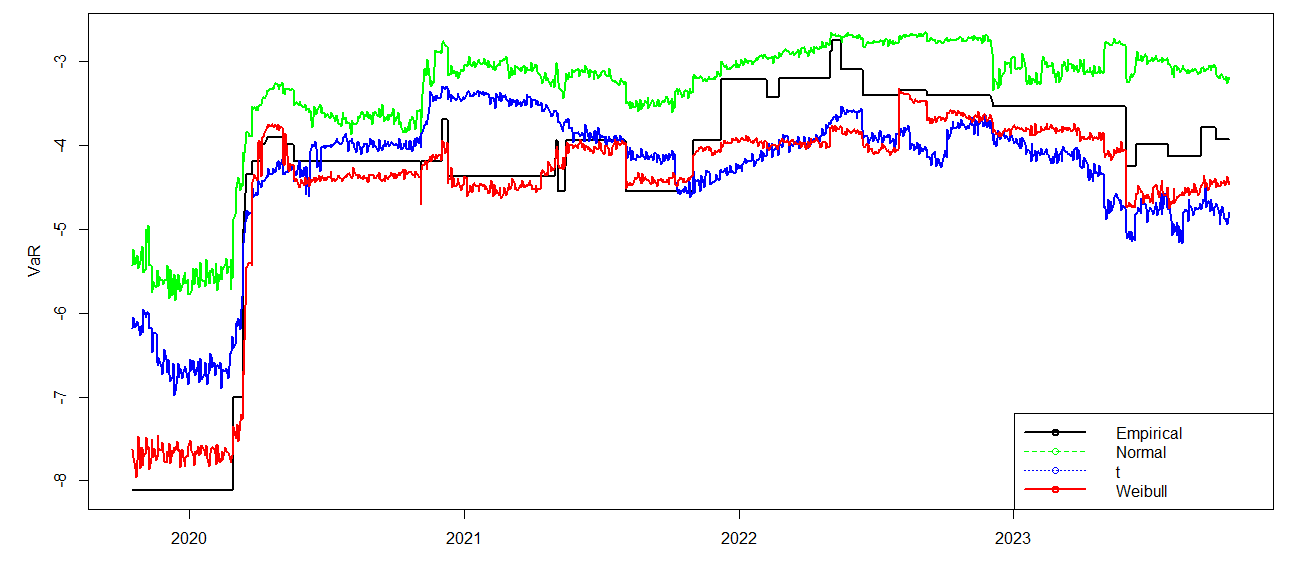}
\caption{One-day 1\% VaR forecasts based on the Gaussian mixture model (green), $t$-mixture model (blue), and mirrored Weibull mixture model (red) for the CVS stock from mid-2019 to mid-2024.}
\label{fig:CVS}
\end{figure}

Finally, the MMW model again shows favourable results for one-day forecasts at 1\% confidence level compared to GMM and $t$MM. It can be observed from Figure \ref{fig:CVS} that the MMW VaR estimates (red line) resemble more closely the empirical estimates (black line) for the majority of the 4 year period between mid-2020 and mid-2024. Numerical results concur that the MMW model is markedly preferable to GMM and $t$MM, with MMW obtaining a much lower average mean square error (MSE) of 0.18 (compared to GMM=1.21 and $t$MM=0.60) and passing Kupiec's test for all cases (failure rate for MMW=0.00, GMM=0.25, and $t$MM=0.03).   

\section{Conclusions}
\label{sec:concl}

We proposed the mixtures of mirrored Weibull (MMW) distributions for modelling financial returns. The MMW model not only possesses nice theoretical properties, such as simple closed-form expression for the density and the cdf, but it also produces more accurate and reliable VaR estimates/forecasts compared to the Gaussian mixture model and the $t$-mixture model when applied to three S\&P500 stocks in Section \ref{sec:data}. 
It is noted that the mirrored Weibull distribution is defined on a semi-infinite interval. Future work could explore modifications of the mirrored Weibull distribution to enable the modelling of data across the the entire real line.

\section*{Acknowledgment}

The work of the first author was supported by the China Scholarship Council (No. 202208845001) and the Inner Mongolia Agricultural University Basic Science Research Initiation Fund Project (No. JC2021002). 



\end{document}